\documentclass[10pt, a4paper, twocolumn, teaser, showabstract]{naverlabseurope}

\usepackage{lipsum}
\usepackage{multicol}
\usepackage{tikz,tkz-kiviat,pgfplots}
\usepackage{tabularx}
\usepackage{booktabs}
\usepackage{amssymb}
\usepackage{amsmath}
\usepackage{float}
\usepackage[utf8]{inputenc}
\usepackage{microtype}
\usepackage{graphicx}
\usepackage{hyperref}

\usepackage{natbib}

\title{XProvence: Zero-Cost Multilingual Context Pruning for Retrieval-Augmented Generation}

\correspondingauthor{youssef.mohamed@kaust.edu.sa,  nadia.chirkova@naverlabs.com}

\authors{Youssef Mohamed$^{1\dagger}$ \authsep Mohamed Elhoseiny$^{1}$ \authsep Thibault Formal$^{2}$ \authsep Nadezhda Chirkova$^{2}$}
\affiliations{$^{1}$KAUST $^{2}$NAVER LABS Europe}
\contributions{$^{\dagger}$ Work done while being an intern at Naver Labs Europe.}
\website{https://huggingface.co/naver/xprovence-reranker-bgem3-v2}
\websiteref{\href{https://huggingface.co/naver/xprovence-reranker-bgem3-v2}}

\teaserfig{\includegraphics[width=\textwidth]{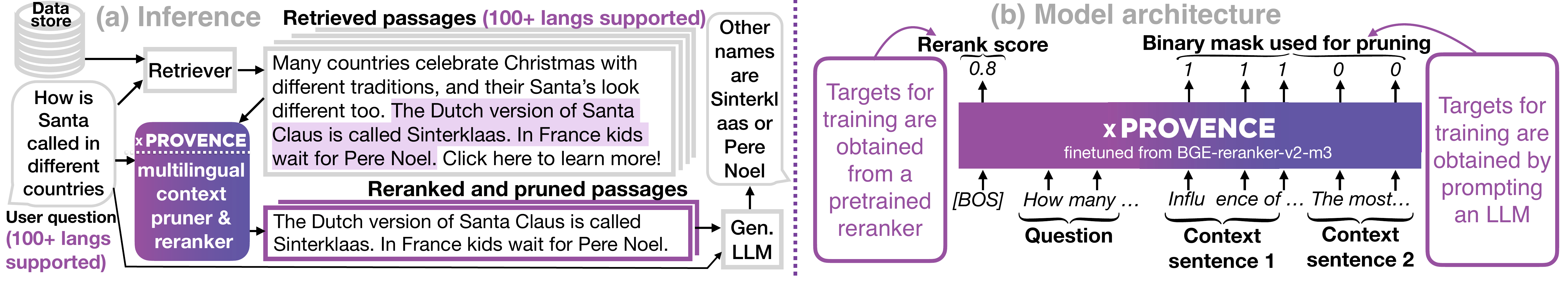}}
\teasercaption{XProvence speeds up generation in  multilingual RAG pipelines through zero-cost context pruning, using an extra prediction head on a multilingual reranker to classify each sentence as relevant or irrelevant to the query. \label{fig:ill}}

\begin{abstract}
This paper introduces XProvence, a multilingual zero-cost context pruning model for Retrieval-Augmented Generation (RAG), trained on 16 languages and supporting 100+ languages through effective cross-lingual transfer. Motivated by the growing use of RAG systems across diverse languages, we explore several strategies to generalize the Provence framework—which first integrated efficient zero-cost context pruning directly into the re-ranking model—beyond English.
 Across four multilingual Question Answering benchmarks, we show how XProvence can prune RAG contexts with minimal-to-no performance degradation and outperforms strong baselines.
 Our training code is available at \url{https://github.com/naver/bergen/tree/main/scripts/xprovence}.

\end{abstract}

\begin{document}

\maketitle

\section{Introduction}
\label{sec:intro}

Retrieval-Augmented Generation (RAG) has emerged as a powerful paradigm for grounding Large Language Models (LLMs) in external knowledge~\citep{lewis2020retrieval}. By retrieving and conditioning on domain-specific contexts, RAG systems have demonstrated strong performance across a wide range of applications. The increasing capabilities of LLMs have in turn made RAG pipelines more effective.

However, the benefits of RAG come with significant computational costs. Retrieved documents substantially increase the input context length, leading to quadratic growth in inference time, higher deployment costs, and a larger carbon footprint. 
Consequently, reducing the size of the context fed to the LLM
has become a key research focus. Among various compression strategies~\citep{wang2023learningfiltercontextretrievalaugmented,yoon-etal-2024-compact,cheng2024xrag,louis-etal-2025-pisco,10.1145/3701551.3703527}, the selective removal of irrelevant content from retrieved documents, known as context pruning, has shown particular promise \citep{jiang-etal-2023-llmlingua,xu2023recomp,chirkovaprovence,hwang2024dslr}.

A state-of-the-art approach in this space is Provence \citep{chirkovaprovence}, which introduces a \emph{zero-cost} pruning mechanism integrated directly into the reranking stage of the RAG pipeline. By leveraging the reranker’s query-aware representations, Provence labels sentences as relevant or irrelevant and prunes non-relevant content prior to generation. This simple yet effective design yields significant runtime improvements without compromising performance. However, Provence remains limited to English, constraining its applicability in multilingual settings~\citep{mrag}. In this work, we address these limitations by introducing XProvence, a multilingual extension of Provence. Our contributions can be summarized as follows:
\begin{itemize}
    \item \textbf{Optimal recipe: }We study various strategies for extending the language coverage of Provence, including \textit{(i)} cross-lingual transfer,
  \textit{(ii)} multilingual data annotation, and
  \textit{(iii)} data translation,
 aiming to find the optimal recipe for training XProvence;
    \item \textbf{Multilingual context pruning: }We introduce XProvence, the first multilingual zero-cost context pruning model, supporting 100+ languages.
    \item \textbf{Extensive evaluation on 4 datasets}: We show that XProvence effectively compresses RAG contexts across languages, achieving minimal-to-no degradation in downstream quality and outperforming strong baselines on non-English benchmarks. 
\end{itemize}
Our model is available at \url{https://huggingface.co/naver/xprovence-reranker-bgem3-v2} and our training code and data are available at \url{https://github.com/naver/bergen/tree/main/scripts/xprovence}.

\section{Background}
\label{sec:provence}
\paragraph{Reranking.} Production-ready RAG systems typically consist of three main components: a retriever, which coarsely identifies relevant passages from a datastore; a reranker, which refines the context selection; and a generator LLM, which produces the final output \citep{rau-etal-2024-bergen}. Rerankers score candidate passages by jointly encoding the query and each passage, which is computationally more expensive than bi-encoder retrieval but typically yields more accurate rankings\footnote{By contrast, retrievers encode the query and each passage independently, which allows passages to be pre-computed offline and makes retrieval substantially faster, but generally at the cost of ranking precision.}. While decoder-style LLM rerankers~\citep{zhuang-etal-2024-beyond, qin-etal-2024-large, sun-etal-2023-chatgpt} have recently advanced the state of the art, BERT-style cross-encoder models remain strong baselines and are often preferred in practice due to their efficiency and ease of deployment. Cross-encoders take as input a concatenation of a \verb|<BOS>| token, a tokenized  query, a \verb|<SEP>| token, and a tokenized passage \citep{nogueira2020passagererankingbert}. A linear head applied to the representation of the \verb|<BOS>| token outputs a relevance score used to rank passages. Rerankers are typically trained on query–document pairs to optimize the relevance of retrieved passages.

\paragraph{Provence.} 
Provence \citep{chirkovaprovence} proposes to enhance rerankers with context pruning capabilities by applying an additional linear head on top of the representations of the \textit{passage tokens} as shown in Figure~\ref{fig:ill} (b). 
A linear head outputs per-token values in the range $(0, 1)$, which are then binarized into $\{0, 1\}$ using a predefined pruning threshold. Each sentence in a passage is then classified as relevant if and only if more than $50\%$ of its tokens are relevant. 
The irrelevant sentences are removed from the passage before passing it to the generator LLM, which results in reduced context lengths and subsequent speed-ups in generation. By integrating context pruning directly into the reranker, the approach introduces no additional cost to a standard RAG pipeline.

Training a Provence model involves fine-tuning a pretrained reranker with a context pruning objective—formulated as per-token binary classification using cross-entropy loss—and a regularization term that preserves reranking capabilities through a regression objective with mean squared error loss. 
Targets for reranking are simply obtained from the pretrained reranker, and targets for context pruning are obtained by prompting a strong LLM, e.g., Llama-3-8B \citep{llama3modelcard}. In particular, an LLM is provided with a query and a passage, and prompted to answer the query using only information in the passage, while citing sentences used in the answer with a \verb|[i]| template.

Prior context pruning approaches such as RECOMP~\citep{xu2023recomp} or DSLR~\citep{hwang2024dslr} encode sentences in a passage independently of each other. In contrast, Provence encodes all the sentences in a retrieved passage together with a query, in a single reranker forward pass. It makes context pruning zero-cost in the RAG pipeline and enables more precise context pruning, since information about coreferences between sentences can now be used to make decisions about sentence pruning. Collectively, these properties establish Provence as an \textit{effective and efficient practical solution for context pruning}. However, it remains limited to English, constraining its applicability in multilingual RAG settings.

\section{Training Methodology for XProvence}
\label{sec:method}
We aim to extend Provence beyond English, enabling effective zero-cost pruning across multiple languages. As a foundation for our XProvence model, we rely on the BGE-M3 reranker\footnote{In particular, we use \texttt{BAAI/bge-reranker-v2-m3}.} \citep{chen2024bge}  which supports 100+ languages. In search of the optimal training recipe, we empirically compare three strategies described below.

\paragraph{\textcolor[HTML]{117733}{\textbf{Cross-Lingual Transfer.}}}
The simplest way to train XProvence is to replicate the original Provence training procedure—using the same English training data—but initialize it from a multilingual reranker.
The multilingual context pruning capabilities of the final model emerge from cross-lingual transfer~\citep{commoncrawl,artetxe-etal-2020-cross,wu-dredze-2019-beto} and the extensive multilingual pretraining of the base model. 
In our experiments, we rely on the MS MARCO-based dataset~\citep{bajaj2018msmarcohumangenerated} available in the Provence repository\footnote{\url{ https://github.com/naver/bergen/tree/main/scripts/provence}}.

\paragraph{\textcolor[HTML]{C92F4B}{\textbf{Data Translation.}}} An alternative strategy, commonly used to extend NLP models beyond English is to translate the English data into multiple languages \citep{bonifacio2022mmarcomultilingualversionms}. 
For each language, we translate 125$k$ randomly sampled query-context pairs from the MS MARCO-based dataset used to train Provence. 
We perform sentence-by-sentence translation into 16 languages using a strong translator LLM, namely GemmaX2 9B 
\citep{cui-etal-2025-multilingual}. 

\paragraph{\textcolor[HTML]{D8A547}{\textbf{Multilingual Data Annotation.}}}
A third strategy is to rely on multilingual-by-design data.
For this purpose, we use the MIRACL dataset \citep{zhang-etal-2023-miracl},  commonly employed to train multilingual retrievers and rerankers.
MIRACL contains queries 
in 16 languages, paired with passages in the same languages sourced from Wikipedia. 
We utilize the multilingual Aya Expanse 8B 
\citep{dang2024ayaexpansecombiningresearch} model to generate synthetic labeling for context pruning, following the same procedure as in Provence.   
We provide instructions in English and ask the model to answer in the query language, as this strategy worked best in our preliminary experiments.

\section{Experimental Setup}
\label{sec:setup}
\paragraph{Training details.}
We train all our models on 4 Nvidia A100 GPUs, and run evaluations using a single A100 GPU per experiment. Similar to Provence, we train for a total of 5 epochs, setting the learning rate to $1e-5$ and batch size to $64$. We utilize spaCy's multilingual sentence tokenizer to split passages into sentences.
Translated and annotated datasets
include 16 languages\footnote{Training data languages: \texttt{ar}, \texttt{bn}, \texttt{en}, \texttt{es}, \texttt{fa}, \texttt{fi}, \texttt{fr}, \texttt{hi}, \texttt{id}, \texttt{ja}, \texttt{ko}, \texttt{ru}, \texttt{sw}, \texttt{te}, \texttt{th}, \texttt{zh}.}. 

\paragraph{Evaluation.} We use the Bergen library~\citep{rau-etal-2024-bergen} for evaluation, with Aya Expanse-8B~\citep{dang2024ayaexpansecombiningresearch} as a generator LLM for all datasets. We run generation with greedy decoding and only evaluate on 23 languages supported by Aya Expanse 8B.

\paragraph{Evaluation datasets.} We conduct evaluations on standard Multilingual Question Answering (MQA) benchmarks: MKQA~\citep{longpre-etal-2021-mkqa} (16 langs\footnote{Considered MKQA languages: seen: \texttt{ar},  \texttt{en}, \texttt{es}, \texttt{fr}, \texttt{ja}, \texttt{ko}, \texttt{ru}, \texttt{zh}; unseen: \texttt{de}, \texttt{he},  \texttt{it}, \texttt{nl}, \texttt{pl}, \texttt{pt}, \texttt{tr}, \texttt{vi}}, 2.8$k$ ex./lang.) and TydiQA~\citep{clark-etal-2020-tydi} (5 langs\footnote{Considered TydiQA languages: \texttt{ar}, \texttt{en}, \texttt{id}, \texttt{ko}, \texttt{ru}}, 400-700 ex./lang.), both relying on a Wikipedia datastore (passages in the same language as the query language or in English). We follow the RAG experimental setup of the Bergen library~\citep{mrag,rau-etal-2024-bergen}, employing both the BGE-M3 multilingual retriever and reranker \citep{chen2024bge} (top-5 passages per query are fed to the LLM), along with the character 3-gram evaluation metric~\citep{mrag}. This metric measures the proportion of character 3-grams from the short ground-truth label that are present in the LLM-generated answer.

To evaluate XProvence beyond the Wikipedia domain, we further consider two other MQA datasets: MedExpQA~\citep{alonso2024medexpqa} (multiple choice medical questions, 4 languages\footnote{MedExpQA languages: \texttt{en}, \texttt{es}, \texttt{fr}, \texttt{it}}, 125 ex./lang.) and XPQA~\citep{shen-etal-2023-xpqa} (questions about e-commerce products, 11 languages\footnote{Considered XPQA languages: \texttt{ar}, \texttt{de}, \texttt{es}, \texttt{fr}, \texttt{hi}, \texttt{it}, \texttt{ja}, \texttt{ko}, \texttt{pl}, \texttt{pt}, \texttt{zh}}, 1.2$k$-1.9$k$ ex./lang.). Both datasets provide a gold context for each query—retrieval is thus not needed. We evaluate MedExpQA using accuracy and XPQA using LLM-as-a-judge~\citep{rau-etal-2024-bergen}.

\paragraph{Baselines.}
We compare training strategies described in Section~\ref{sec:method}. We also include DSLR—the most effective baseline from the Provence paper~\citep{hwang2024dslr}. 
\textcolor[HTML]{999999}{\textbf{DSLR}} segments input contexts into individual sentences and encodes each sentence together with the query using a pretrained reranker.
Sentences whose reranking scores exceed a predefined threshold are retained in their original order.
However, DSLR faces a key limitation addressed by XProvence: it encodes sentences independently, thereby neglecting semantic relationships across sentences.
For a fair comparison, we evaluate DSLR using the BGE-M3 reranker.

\begin{figure*}[t]
    \centering
    \includegraphics[width=0.98\linewidth]{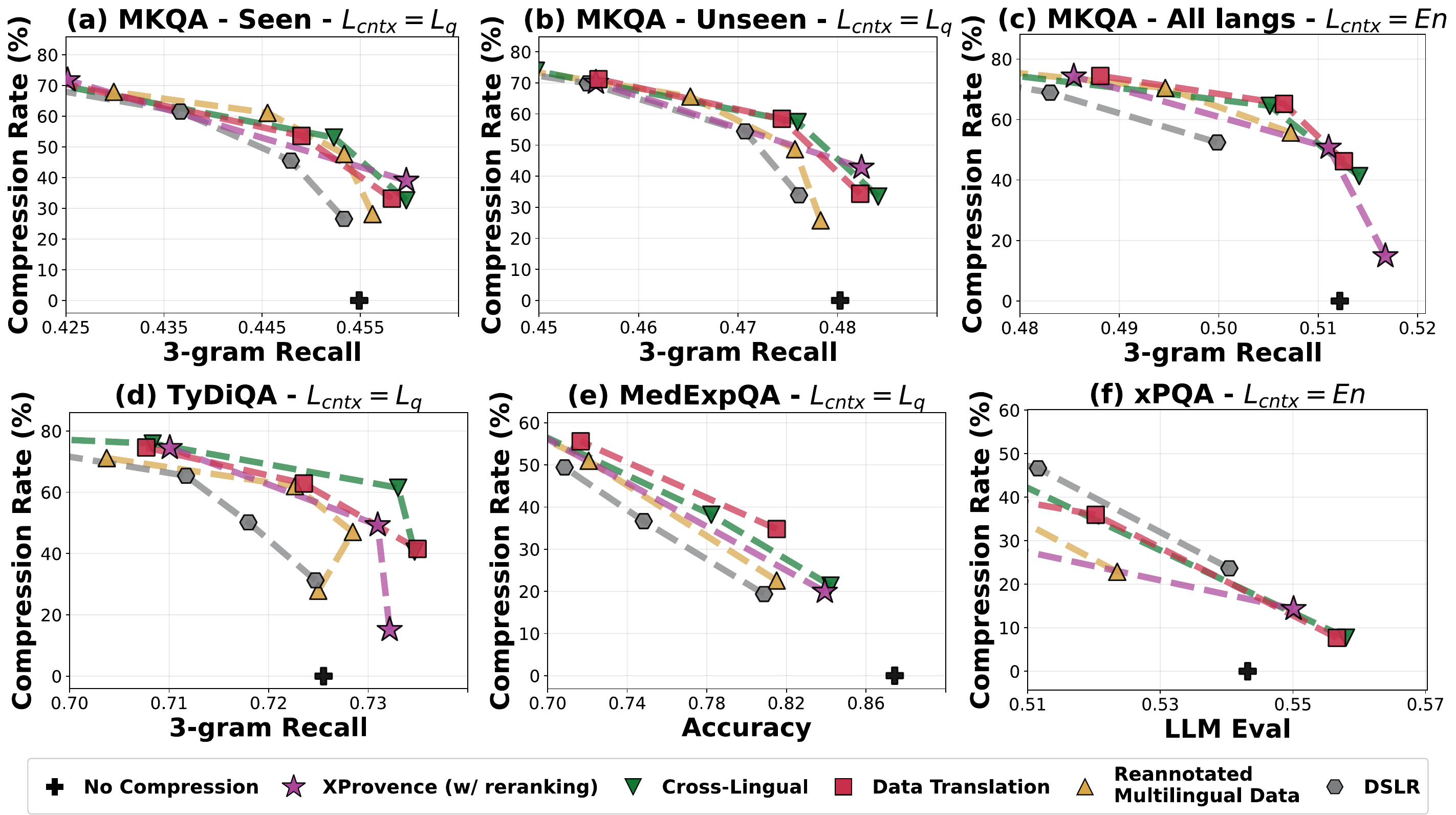}
    \caption{
    Main results: for each dataset, we present a Pareto front obtained by varying the pruning threshold and averaged over languages. Lines located closer to the top right corner are best performing. Notation $L_{q}$/$L_{cntx}$ denotes the language of query/passage. For MKQA, we perform controlled experiments with various language settings, including  settings $L_{cntx}=L_{q}$ and $L_{cntx}=En$. For the former setting,  we present separately the results for languages seen and unseen in the training data. For the remaining datasets, we follow their original setting.
    }
    \label{fig:results}
\end{figure*}

\section{Experiments}
\label{sec:experiments}

Results are shown in Fig.~\ref{fig:results}\footnote{Per-language results are provided in the code repository.}. For each dataset, we plot the task metric on the $x$ axis, the context compression rate\footnote{Context compression is defined as the average portion of pruned-out context.} on the $y$ axis, and present a Pareto front obtained by varying the pruning threshold. 
Lines located closer to the top right corner are best performing. We answer several research questions in the following.

\begin{table*}[t!]
    \centering
    \begin{tabular}{lcc}
     \toprule
        Reranker & MKQA (R@20) & MIRACL (nDCG@10) \\ \midrule
        BGE-M3   & 70.3 & 74.5 \\
        XProvence (ours) & 70.7 & 74.3 \\
       \bottomrule
    \end{tabular}
    \caption{Reranking Performance averaged across languages. 
    }
    \label{tab:reranking}
\end{table*}

\paragraph{RQ1: What is the best training recipe for XProvence?} 
We compare the three training strategies described in Section~\ref{sec:method}: \textcolor[HTML]{117733}{\textbf{Cross-Lingual Transfer (CLT)}} \textit{vs} \textcolor[HTML]{C92F4B}{\textbf{Data Translation (DT)}} \textit{vs} \textcolor[HTML]{D8A547}{\textbf{Multilingual Data Annotation (MDT)}}. 
For simplicity, all models in this subsection are trained for context compression only, without reranking loss.

Our conclusions are as follows: {\it (1)} The \textcolor[HTML]{117733}{\textbf{CLT}} strategy is surprisingly effective and reaches similar or superior performance compared to other strategies on all datasets; {\it(2)} Comparing \textcolor[HTML]{C92F4B}{\textbf{DT}} {\it vs} \textcolor[HTML]{117733}{\textbf{CLT}}, we observe that translating MS MARCO data does not bring consistent improvements; {\it(3)} Comparing \textcolor[HTML]{D8A547}{\textbf{MDT}} {\it vs} \textcolor[HTML]{117733}{\textbf{CLT}}, we observe that training on multilingual by-design but limited-domain data (Wikipedia-based MIRACL) leads to lower performance compared to training on the more diverse MS MARCO data, i.e., the \textcolor[HTML]{117733}{\textbf{CLT}} lines are located closer to the top right corners on Pareto fronts. 

For our final model, \textcolor[HTML]{AA4499}{\textbf{XProvence (w/reranking)}}, we use the data translation strategy trained with Provence’s joint objective. Our results on the effectiveness of cross-lingual transfer align well with the prior results reported in the literature~\citep{wu-dredze-2019-beto,pires-etal-2019-multilingual,K2020Cross-Lingual}. 

\paragraph{RQ2: How does XProvence compare against DSLR?} Comparing our final model, \textcolor[HTML]{AA4499}{\textbf{XProvence (w/ reranking)}} and \textcolor[HTML]{7B7D7F}{\textbf{DSLR}}, we observe that \textcolor[HTML]{AA4499}{\textbf{XProvence}} outperforms \textcolor[HTML]{7B7D7F}{\textbf{DSLR}} in all cases, except XPQA. We note that \textcolor[HTML]{AA4499}{\textbf{XProvence}} is zero-cost, while \textcolor[HTML]{7B7D7F}{\textbf{DSLR}} incurs extra computational costs in the RAG pipeline, 
as reranking and context pruning require separate forward passes.

\paragraph{RQ3: Does XProvence enable context compression with minimal-to-no performance loss?} The pruning strength of XProvence can be adjusted through a pruning threshold. We observe that for MKQA and TyDiQA, \textcolor[HTML]{AA4499}{\textbf{XProvence}} enables the compression rate of 40—60\%, 
with the same performance as
the full context setting.
For XPQA and MedExpQA, the achieved pruning compression is lower, due to the nature of gold contexts which are provided as parts of the datasets. 

\paragraph{RQ4: Does XProvence perform well on languages unseen in the training data and when the context language differs from the query language?} We consider these settings for MKQA in Figure~\ref{fig:results} (b) and (c) correspondingly and find that the performance of \textcolor[HTML]{AA4499}{\textbf{XProvence}} is robust to such settings. This is the result of the effective cross-lingual transfer discussed above.

\paragraph{RQ5: Does integration of context pruning preserve reranking performance?}

We evaluate in Table~\ref{tab:reranking} the reranking performance of XProvence using FlagEmbedding\footnote{https://github.com/FlagOpen/FlagEmbedding} with default settings. We show that XProvence maintains the reranking performance of the base model, while being able to perform context pruning.

\section{Conclusion}
In this work, we present XProvence, a reranking model equipped with integrated context pruning, capable of handling queries and contexts in over 100+ languages. Through an empirical investigation of training strategies on multiple MQA benchmarks, we show the effectiveness of an easy-to-implement strategy, consisting of tuning a multilingual reranker on English context-pruning data and exploiting cross-lingual transfer. 

\label{sec:conclusion}

    \bibliographystyle{iclr2025_conference}
    \bibliography{custom}

\end{document}